\documentclass[a4 paper]{article}
\usepackage{epsfig}
\usepackage{multicol}
\usepackage{array}
\usepackage{amsfonts,amsmath,amssymb}

\input{epsf.sty}
\begin{document}
\twocolumn[
\begin{center}
{\Large{\bf The generalized Milne problem in gas-dusty atmosphere}}\\
\bigskip
{N. A. Silant'ev\thanks{E-mail: nsilant@bk.ru}\,\,, G. A. Alekseeva,\, V. V. Novikov}
\medskip\\
\end{center}
\begin{center}
 {Central Astronomical Observatory at Pulkovo of Russian Academy of Sciences,\\ 196140,
Saint-Petersburg, Pulkovskoe shosse 65, Russia\\}
\medskip
\end{center}

\bigskip
\begin{center}
{received..... 2017, \qquad accepted....}
\end{center}]
\begin{abstract}
 We consider  the generalized Milne problem in non-conservative plane-parallel optically thick atmosphere consisting of
two components - the free electrons and small dust particles. Recall, that the traditional Milne problem  describes the propagation of
radiation through the conservative (without absorption) optically thick atmosphere when the source of thermal radiation located far
below the surface. In such case, the flux of propagating light is the same at every distance in an atmosphere. In the generalized Milne problem, the flux  changes  inside the atmosphere. The solutions of the both Milne problems give the 
angular distribution and polarization degree of emerging radiation.
   The considered problem depends on two dimensionless parameters  W and (a+b), which depend on three parameters: $\eta$ - the ratio of optical depth due to free electrons to optical depth due to small dust grains; the absorption factor $\varepsilon$ of
dust grains and two coefficients - $\overline b_1$ and $\overline b_2$, describing the averaged anisotropic dust grains. These coefficients
obey the relation $\overline b_1+3\overline b_2=1$. The goal of the paper is  to study the dependence of the radiation angular distribution 
and degree of polarization of emerging light on these  parameters.  Here we consider only continuum radiation.

\end{abstract}

\section{Introduction}

One of important problems in radiative transfer theory is the Milne problem. This problem is related with the solution of the transfer equation when the sources of non-polarized radiation are located far from the boundary of optically thick atmosphere. The most known
example is the usual Milne problem for free electron non-absorbed atmosphere. This problem was solved by Chandrasekhar (1960). His results for the angular distribution $J(\mu)=I(\mu)/I(0)$ and polarization degree $p(\mu)$ of the emerging radiation are used in many applications.
Recall, that $\mu=\cos\vartheta$, where $\vartheta$ is the angle between line of sight ${\bf n}$ and the outer normal ${\bf N}$ to the plane-parallel optically thick atmosphere.
 The angular distribution $J(\mu)$ increases with the increase of $\mu$ up to value $J(1)=3.06$. The polarization degree
is $p(1)=0$ and inreases up to $11.71\% $ at $\mu=0$. 

The small dust grains with the spherical form and the molecules with isotropic polarizability  $\beta$ scatter the radiation by the same law as the electrons. If the absorption factor of
the dust substance $\varepsilon$ increases,  the angular distribution and polarization degree also increase.  For example, in the case  
$\varepsilon=0.1$ one takes place $J(1)=4.39$ and $p(0)=20.35\%$  (see Silant'ev 1980).

The dust grains and molecules  are characterized by the anisotropic polarizability tensor $\beta_{ij}(\omega)$, where $\omega$ is the cyclic frequency of light.
These scattering particles due to chaotic thermal motions are freely oriented in an atmosphere.

The radiative transfer equation for this case firstly was derived by Chandrasekhar (1960). 
This equation depends on two parameters - $\overline{b}_1$ , describing the Rayleigh scattering, and the additional term $\overline{b}_2$, which describes the effect of anisotropy of scattering particles. This term (depolarization factor) describes the  additional isotropic non-polarized part of scattered radiation.
 The depolarizing effect of anisotropy of particles mostly reveals by consideration of axially symmetric problems.  One of such problems is the Milne problem.

In axially symmetric  problems the radiation is described by two intensities - $I_l(\tau,\mu)$ and $I_r(\tau,\mu)$.
Here $\tau $ is the optical depth below the surface of semi-infinite plane-parallel atmosphere.
 The intensity $I_l$ describes the light linearly polarized in the plane (${\bf nN}$), and $I_r$ is the light intensity with polarization
perpendicular to this plane. The total intensity $I=I_l+I_r$, and the Stokes parameter $Q=I_l-I_r$. The Stokes
parameter $U\equiv 0$. The degree of linear polarization is characterized by  $p=Q/I$. Note that case $ Q<0$ corresponds to the wave electric field oscillations perpendicular to the plane $({\bf nN})$. This case holds in the Milne problem.   Frequently one uses the radiative transfer equation for parameters $I(\tau,\mu)$ and $Q(\tau,\mu)$ .

 The factorization of the matrix phase function $\hat P(\mu,\mu')$, i.e. the presentation  this matrix as a product of two matrices $\hat
P(\mu,\mu')=\hat A(\mu)\hat A^{T}(\mu')$, plays very important role in radiative transfer theory. Factorization is not unique (see Hulst 1980). Factorization used in the papers Lenoble (1970),
Abhyankar \& Fymat (1971) and Schnatz \& Siewert (1971) describes the  matrix phase function for the Rayleigh scattering.
  In Frisch (2017)  the new factorizations for linear combination of Rayleigh and isotropic scatterings are given.  Note that in this private communication the factorizations are given both for $(I,Q)$ and $(I_l, I_r)$ cases. Below we consider only the $(I, Q)$ case.

\section{Radiative transfer equation }

First let us recall the radiative transfer equation for the (column) vector with the components ($I, Q$) in an atmosphere
consisting of averaged small anisotropic particles and free electrons.   The equation  for $I(\tau,\mu)$  and $Q(\tau,\mu)$  can  be readily transformed from the equation for the  column ($I_l(\tau,\mu)$ , $I_r(\tau,\mu)$) presented in Chandrasekhar 1960; Dolginov et al. 1995; Silant'ev et al. 2015 :

\[
\mu \frac{d}{d\tau}\left (\begin {array}{c} I(\tau,\mu) \\ Q(\tau,\mu) \end{array}\right )
=\left (\begin {array}{c} I(\tau,\mu) \\ Q(\tau,\mu) \end{array}\right )-\frac{1}{2}(a+b)\int^1_{-1}d\mu'\times
\]
\[
\left(\begin{array}{rr}1+C(1-3\mu^2)(1-3\mu'^2)\,, \,3C(1-3\mu^2)(1-\mu'^2)\\
 \,3C(1-\mu^2)(1-3\mu'^2)\,, 9C(1-\mu^2)(1-\mu'^2)\end{array}\right )
\]
\begin{equation}
\times \left (\begin {array}{c} I(\tau,\mu) \\ Q(\tau,\mu) \end{array}\right )-s(\tau)\left (\begin {array}{c} 1 \\ 0 \end{array} \right ),
\label{1}
\end{equation}
\noindent where $d\tau=\kappa dz$ determines the dimensionless optical depth. Extinction factor $\kappa=N_g(\sigma^{(s)}_g+\sigma^{(a)}_g)+N_e\sigma_T$. $\sigma^{(s)}_g$ and $\sigma^{(a)}_g$ are the
cross-sections of scattering and absorption by dust grains, $\sigma^{(t)}_g$ is cross-section of total extinction, $\sigma_T$ is the Thomson cross-section. $N_g$ and $N_e$ are the number densities of  grains and free electrons, respectively.  The degree of light absorption
 $\varepsilon=\sigma^{(a)}_g/\sigma^{(t)}_g$, $\mu ={\bf nN}$ is cosine of the angle between the directions of light propagation ${\bf n}$ and the outer normal ${\bf N}$ to plane-parallel semi-infinite atmosphere.   The parameter C=W/8. The parameters $a$ , $b$, $\eta$ and $W$ are:
\[
a=\frac{\eta+(1-\varepsilon)\overline{b}_1}{1+\eta},
\]
\begin{equation}
b=\frac{3(1-\varepsilon)\overline{b}_2}{1+\eta},
\label{2}
\end{equation}
\begin{equation}
\eta=\frac{N_e\sigma_T}{N_g\sigma^{(t)}_g},
\label{3}
\end{equation}
\[
W=\frac{a}{a+b}=\frac{\eta+(1-\varepsilon)\overline b_1}{\eta+(1-\varepsilon)},
\]
\begin{equation}
a+b=\frac{\eta+(1-\varepsilon)}{1+\eta}.
\label{4}
\end{equation}
\noindent   The parameters $\overline b_1$ and $\overline b_2$ obey the
relation (8). This relation was used in expressions (4) for $W$ and $(a+b)$. So, Eq.(1) depends on two  parameters -
$W$ and $(a+b)$.

 The value $s(\tau)$ is the source of non-polarized radiation.
 Usual source of radiation is the thermal radiation:
\begin{equation}
s(\tau)=B_{\omega}(T(\tau)),
\label{5}
\end{equation}
\noindent where $B_{\omega}(T)$ is the Planck function with the temperature, depending on the optical depth $\tau$. This
dependence appears in various models of atmospheres.

 For reader's convenience, we shortly explaine the physical sense of parameters $\overline{b}_1$ and $\overline{b}_2$.
The scattering cross-section of small particles (dust grains, molecules) is (Dolginov et al. 1995):
\begin{equation}
\sigma_s=\frac{8\pi}{3}k^4(b_1+3b_2),
\label{6}
\end{equation}
\noindent where $\omega=2\pi\nu$ is cyclic frequency of light, $k=\omega/c$, $c$ is the speed of
light.   The values $b_1$ and $b_2$ are related to polarizability tensor $\beta_{ij}(\omega)$ of
a particle as a whole. Induced dipole moment of a particle, as a whole, is equal to $p_i(\omega)=\beta_{ij}(\omega)E_j(\omega)$.
Anisotropic particle with axial symmetry is characterized by two polarizabilities - along the symmetry axis $\beta_{\parallel}(\omega)$, and in transverse direction $\beta_{\perp}(\omega)$. For such particles:

\[
b_1=\frac{1}{9}|2\beta_{\perp}+\beta_{\parallel}|^2+\frac{1}{45}|\beta_{\parallel}-\beta_{\perp}|^2,
\]
\begin{equation}
b_2=\frac{1}{15}|\beta_{\parallel}-\beta_{\perp}|^2.
\label{7}
\end{equation}
\noindent In transfer equation we use the dimensionless parameters

\[
\overline{b}_1=\frac{b_1}{b_1+3b_2},\,\,\, \overline{b}_2=\frac{b_2}{b_1+3b_2},
\]
\begin{equation}
\overline{b}_1+3\overline{b}_2=1.
\label{8}
\end{equation}
\noindent For needle like particles ($|\beta_{\parallel}|\gg |\beta_{\perp}|)$ parameters $\overline{b}_1=0.4,\,
\overline{b}_2=0.2$, and for plate like particles ($|\beta_{\perp}|\gg |\beta_{\parallel}|$) we have
$\overline{b}_1=0.7,\, \overline{b}_2=0.1$.
Parameter $\overline{b}_2$ describes the depolarization of radiation, scattered by freely oriented anisotropic
particles. So, the needle like particles depolarize radiation greater than the plate like ones.

 The absorption of light is described by imaginary part
of the polarizability of freely oriented grain: $\sigma^{(a)}_{grain}=4\pi\,k\,Im\langle \beta(\omega)\rangle$ (see Fr\"olich 1958, Dolginov et al. 1995). The estimate of polarizability $\beta$ may be taken from the formula $\beta=a^3(m^2-1)/(m^2+2)$ , where $a$ is radius of the  dust grain and $m$ is refraction index of the dust subject ($m=m'+im''$). If $m$ is large, the polarizability $\beta\simeq a^3$.  For the graphite grains Greenberg (1968) gives $m'=2.45$ and $m''\simeq 1.45$ ( for wavelength $\lambda\simeq 0.5 $ mk.)

 The  matrix phase function in Eq.(1) can be written as the product $\hat A(\mu)\hat A^T(\mu')$  (Frisch 2017). The  matrix $\hat A(\mu)$ is equal to:
\begin{equation}
\hat A(\mu)= \sqrt{(a+b)}\left (\begin{array}{rr}1 ,\,\, \,\sqrt{C}(1-3\mu^2) \\ 0 ,\,\,\,3\sqrt{C} (1-\mu^2) \end{array}\right).
\label{9}
\end{equation}
\noindent  The superscript $T$ stands for the matrix transpose.  We emphasize that the new factorization (9) describes the linear combination of Rayleigh and isotropic light scattering.

It is useful  introduce the vector ${\bf K}(\tau)$:
\begin{equation}
{\bf K}(\tau)=\frac{1}{2}\int_{-1}^1\,d\mu\hat A^T(\mu){\bf I}(\tau,\mu).
\label{10}
\end{equation}
\noindent Using this notion,  Eq.(1)  can be written in the form:
 \begin{equation}
\mu \frac{d{\bf I}(\tau,\mu)}{d\tau}= {\bf I}(\tau,\mu)- \hat{A}(\mu){\bf K}(\tau) -s(\tau)\left (\begin {array}{c} 1 \\ 0 \end{array} \right ),
\label{11}
\end{equation}
\noindent where  the matrix $\hat{A}(\mu)$ is given in Eq.(9).

It is easy verify that Eq.(1) gives rise to  the  conservation law of radiative flux. 
Taking $\varepsilon=0$ and $s(\tau)=0$, we obtain:

\begin{equation}
\frac{dF(\tau)}{d\tau}=0, \,\,\,F(\tau)=\int_{-1}^{1}d\mu\,\mu\,I(\mu,\tau).
\label{12}
\end{equation}

\subsection{ The solution by resolvent technique}

The general theory to calculate the vector ${\bf K(\tau)}$ is  presented in Silant'ev et al. (2015).
   Recall, that according to this theory the vector ${\bf K}(\tau)$  obeys the integral equation, which has solution through the resolvent matrix  $\hat{R}(\tau,\tau')$. This matrix can be known if we know the matrices $\hat{R}(\tau,0)$ and $\hat{R}(0,\tau)$.
The kernel $\hat{L}(|\tau-\tau'|)$ of equation for $\hat{R}(\tau,\tau')$ is symmetric: $\hat{L}=\hat{L}^T$ . This gives rise to the relation  $\hat{R}(\tau,\tau')=\hat{R}^T(\tau',\tau)$. The Laplace transform of $R(\tau,0)$ over parameter $1/\mu$ is known  as $H(\mu)$- matrix.
This matrix obeys the following  nonlinear equation:

\begin{equation}
\hat{H}(\mu)=\hat{E} +\mu\int_0^1d\mu'\frac{
\hat{H}(\mu)\hat{H}^T(\mu')\hat{\Psi}(\mu')}{\mu+\mu'}.
\label{13}
\end{equation}
\noindent  The martrix $\hat{\Psi}(\mu)$ is related with $\hat{A}(\mu)$:
\begin{equation}
\hat{\Psi}(\mu)=\frac{1}{2}\hat{A}^T(\mu)\hat{A}(\mu).
\label{14}
\end{equation}

The linear equation for matrix   $\hat H(\mu)$ is the following:
\[
\left(\hat{E}-2\int_0^1 d\mu\frac{\hat{\Psi}(\mu)}{1- k^2\mu^2}\right)\hat{H}\left(\frac{1}{k}\right)=
\]
\begin{equation}
\left(\hat{E} -\int_0^1 d\mu\frac{
\hat{\Psi}(\mu)\hat{H}(\mu)}{1- k \mu}\right),
\label{15}
\end{equation}
\noindent where $k$ is an arbitrary number.

\section{Formulas for the Milne problem}

The specific feature of the Milne problem is that we are to solve integral equation for ${\bf K}(\tau)$  without the free term (see Sobolev 1969).    The vector ${\bf I}(0,\mu)$, describing the emerging radiation, has the form:

\[
{\bf I}(0,\mu)=\hat{A}(\mu)\int_0^{\infty}\frac{d\tau }{\mu}\exp{\left (-\frac{\tau}{\mu}\right )}{\bf K}(\tau)\equiv
\]
\begin{equation}
\frac{1}{\mu}\hat{A}(\mu)\tilde{{\bf  K}}\left (\frac{1}{\mu}\right ),
\label{16}
\end{equation}
\noindent  i.e. this expression is  proportional to the Laplace transform of ${\bf K}(\tau)$  over variable $ \tau $.
The homogeneous equation for ${\bf K}(\tau)$ without the source term has the form:
\begin{equation}
{\bf K}(\tau)|_{hom}=\int_0^{\infty} d\tau'\hat L(|\tau-\tau'|)\,{\bf K}(\tau')|_{hom},
\label{17}
\end{equation}
\noindent where the kernel $\hat L(|\tau-\tau'|)$  is the following:
\begin{equation}
\hat L(|\tau-\tau'|)=\int^1_0\frac{d\mu}{\mu}\exp{\left(-\frac{|\tau-\tau'|}{\mu}\right)}\hat \Psi(\mu).
\label{18}
\end{equation}

 Further we follow to simple approach  by Sobolev (1969), generalizing his method for the  vector
case. For simplicity we omit the subscript $| hom $.  According to Eqs.(17) and (18), we derive the value ${\bf K}(0)$:
 \[
{\bf K}(0)=\int_0^{\infty} d\tau\hat L(\tau)\,{\bf K}(\tau')\equiv
\]
\begin{equation}
\int_0^1\frac{d\mu}{\mu}\hat {\Psi}(\mu)\tilde {\bf K}\left(\frac{1}{\mu}\right).
\label{19}
\end{equation} 
\noindent Let us get the equation for derivative $d{\bf K}(\tau)/d\tau$, taking into account that the kernel $\hat L(|\tau-\tau'|) $ depends
on the difference $ \tau-\tau' $:
\[
\frac{d{\bf K}(\tau)}{d\tau}=\hat L(\tau){\bf K}(0)+
\]
\begin{equation}
\int_0^{\infty} d\tau'\hat L(|\tau-\tau'|)\,\frac{{\bf K}(\tau')}{d\tau'}.
\label{20}
\end{equation}
 The general solution of Eq.(20) consists of the sum of two terms - the nonzero solution of homogeneous Eq.(17) with some constant $ k $ , i.e. $ k{\bf K}(\tau)$ , and the
solution of non-homogeneous equation (20). The latter is proportional to ${\bf K}(0)$ with some factor. This factor  obeys the 
equation  for $\hat R(\tau,0)$, i.e. the general solution of Eq.(20) has the form:

\begin{equation}
\frac{d{\bf K}(\tau)}{d\tau}=k{\bf K}(\tau) +\hat {R}(\tau,0){\bf K}(0).
\label{21}
\end{equation}
Now let us derive the Laplace transform of this equation. The Laplace transform of the left part of this equation is equal to:
\[
\int_0^{\infty}d\tau \exp{\left(-\frac{\tau}{\mu}\right)}\frac{d{\bf K}(\tau)}{d\tau}=
\]
\begin{equation}
-{\bf K}(0)+\frac{1}{\mu}\tilde{\bf K}\left(\frac{1}{\mu}\right).
\label{22}
\end{equation}
\noindent The Laplace transform of the right part of Eq.(21) has the form:
\begin{equation}
k \tilde{{\bf K}}\left(\frac{1}{\mu}\right)+\tilde{\hat R}\left(\frac{1}{\mu},0\right){\bf K}(0).
\label{23}
\end{equation}
\noindent The equality of Eq.(22) with  Eq.(23)  gives rise to the relation:
\begin{equation}
\frac{1}{\mu}\tilde {\bf K}\left(\frac{1}{\mu}\right )=\frac{\hat H(\mu){\bf K}(0)}{1-k\mu}.
\label{24}
\end{equation}
\noindent Here we  used the relation $\tilde{\hat R}(1/\mu,0) =(\hat H(\mu)- \hat E)$ ( see Silant'ev et al.(2015)).  $\hat E$ is unit matrix.
  Substituting this formula into Eq.(16), we obtain the expression for ${\bf I}(0,\mu)$:
\begin{equation}
{\bf I}(0,\mu)=\frac{A(\mu)\hat H(\mu){\bf K}(0)}{1-k\mu}\equiv \frac{\hat D(\mu){\bf K}(0)}{1-k\mu}.
\label{25}
\end{equation}
\noindent Here we introduced the new matrix:
\begin{equation}
\hat D(\mu)=\hat{A}(\mu)\hat H(\mu).
\label{26}
\end{equation}
\noindent Using Eq.(13),  we obtain  the following equation for matrix $\hat D(\mu)$:

\[
\hat{D}(\mu)\equiv \left (\begin{array}{rr}h(\mu) ,\,\, g(\mu) \\ e(\mu),\,\,\,f(\mu) \end{array}\right)=\hat{A}(\mu)+
\]
\begin{equation}
\frac{\mu}{2} \int_0^1\frac{d\mu'}{\mu+\mu'}\hat{D}(\mu)\hat{D}^T(\mu')\hat{A}(\mu').
\label{27}
\end{equation}
\noindent  The kernel in Eq.(27)  does not depend on $\mu^4$, i.e. this equation is simpler than Eq. (13).

The value ${\bf I}(0,\mu)$ is proportional to ${\bf K}(0)$. According to Eq.(19) the value  ${\bf K}(0)$   related with expression (24).
As a result, we obtain the homogeneous equation for  ${\bf K}(0)$:
\begin{equation}
\left( \hat E-\int_0^1\,d\mu \frac{\hat {\Psi}(\mu)\hat H(\mu)}{1-k\mu}\right){\bf K}(0)=0,
\label{28}
\end{equation}
\noindent or in new matrix $\hat D(\mu)$
\begin{equation}
\left( \hat E-\frac{1}{2}\int_0^1\,d\mu \frac{\hat A^T(\mu)\hat D(\mu)}{1-k\mu}\right){\bf K}(0)=0.
\label{29}
\end{equation}

 This homogeneous equation allows us to obtain only the ratio $K_1(0)/K_2(0)$. So, the expression ${\bf I}(0,\mu)$  contains an arbitrary Const. This Const can be expressed through the observed flux of outgoung radiation. Note that the angular distribution $J(\mu)=I(0,\mu)/I(0,0)$ and the degree of polarization $p(\mu)=Q(0,\mu)/(I(0,\mu)$ are independent of Const.  The negative $p(\mu)$ denotes that the wave electric field oscillations are perpendicular to the plane ${\bf (nN)}$.

The necessary condition to obtain ${\bf K}(0)$ is the zero of the determinant of expression (28), which is the right part of Eq.(15).  Thus,  from Eq.(15) we obtain the separate  equation for the characteristic number $k$:
\begin{equation}
\left|\left(\hat{E}-2\int_0^1 d\mu\frac{\hat{\Psi}(\mu)}{1- k^2\mu^2}\right)\right|=0.
\label{30}
\end{equation}
\noindent This equation is not related with the calculation of the matrix $\hat H(\mu)$.
 It is easy verify that the solution of Eq.(30) for $\varepsilon=0$ gives $k=0$.

  For $\varepsilon<<1$ we obtain  the approximation formula:
\begin{equation}
k_{appr}=\sqrt{\frac{3\varepsilon}{1+\eta}}.
\label{31}
\end{equation}
\noindent The values $k_{appr}$ are given in Tables 1 and 2. It is of interest that Eq.(31) is independent of parameters $\overline{b}_1$ and $\overline{b}_2$,  which characterize the form of grains.

    Let us shortly discuss the calculation of angular distribution $J(\mu)$ and degree of polarization $p(\mu)$.     For the cases $\varepsilon\neq 0$, we have used the direct iterations of  matrix equation (27). But for $\varepsilon=0$  this technique gives the relative error $\sim 5\%$. This is well-known problem beginning from the Chandrasekhar's works. It is related with very slow
convergence of iterations at $\varepsilon=0$.

 To solve this problem Chandrasekhar presented the equation for $H$-function in the form of continuous fraction. In the case of matrix equation Silant'ev (2007) used the same method but for one component of the matrix, hoping that the fast convergence of this equation 
 results in fast convergence of other components. 

 The equation (27)  for $\varepsilon=0$ ($a+b=1$)  gives the following equation for $h(\mu)$:
\begin{equation}
h(\mu)=1+\frac{1}{2} \mu \int_0^1\frac{d\mu'}{\mu+\mu'}[h(\mu)h(\mu')+g(\mu)g(\mu')].
\label{32}
\end{equation}
\noindent The zero's moments $h_0$ and $g_0$ (recall, that they are simple $\mu$-integrals) are equal to
$h_0=2$ and $g_0=0$. Using the equality $\mu/(\mu+\mu')=1 -\mu'/(\mu+\mu')$ and the values $h_0$ and $g_0$, we derive the
following equation for $h(\mu)$:
\begin{equation}
h(\mu)=\frac{2-g(\mu)\int_0^1\frac{d\mu'\mu'}{\mu+\mu'}g(\mu')}{\int_0^1\frac{d\mu'\mu'}{\mu+\mu'}h(\mu')}.
\label{33}
\end{equation}
\noindent Iteration of this equation is fast convergented continuous fraction. Other equations for $g(\mu)$, $e(\mu)$ and $f(\mu)$ are iterated without modifications. For every iteration we use the semi-sum of two precedingly iterated functions.  Results
 in Table 3, corresponding to $\varepsilon=0$, were obtained in such  a manner.

  For pure absorbing dust grains ($\sigma^{(s)}_{grain}=0,\varepsilon=1$) the parameters $W=1$ , $a+b=\eta/(1+\eta)$.
 In this case the term, describing the light scattering by dust grains,  disappears. The parameter $\eta/(1+\eta)$ in the electron scattering term has the sense of  $(1-\varepsilon_{eff})$ with the effective absorption $\varepsilon_{eff}=1/(1+\eta)$.   So, the pure absorbing dust grains play the role of effective absorption in the electron scattering. Note that $\varepsilon_{eff}\to 0$, if the 
parameter  $\eta\to \infty$.

 The intensities $I(\mu)$ and $Q(\mu)$ depend on two parameters - $(a+b) $ and $W=a/(a+b)$, which depend on $\eta$, $\varepsilon$ and $\overline{b}_1$. The different
relations between them give rise to various forms of angular distribution $J(\mu)$ and degree of polarization $p(\mu)$. In particular,
for the spherical dust grains ($\overline{b}_1=1$, $\overline{b}_2=0$ , i.e. $ W=\eta/(\eta+1-\varepsilon)$) there exists the interesting feature. If the parameters $\eta$ and 
$\varepsilon$ obey the relation $\varepsilon_1/(1+\eta_1)=\varepsilon_2/(1+\eta_2)$, then the intensities $I(\mu;\varepsilon_1,\eta_1)=I(\mu;\varepsilon_2, \eta_2)$ and $Q(\mu;\varepsilon_1,\eta_1)=Q(\mu;\varepsilon_2,\eta_2)$.

\section{Results of calculations}

 The characteristic number $k$ increases with the increasing of the absorption degree  $\varepsilon$. We see from Eq.(25) (see the term 
$(1-k\mu $)  in the denominator) that the increase of $k$ gives rise to greater sharpness of the angular distribution of emerging radiation $J(\mu)\gg J(\mu=0)\equiv1.$  So, the presence of large absorption results in the peak like emerging intensity along the normal ${\bf N}$.

 The polarization of emerging  radiation at $\mu \sim 1$   is small. This holds due to symmetry of the problem.  Recall, that the scattering of the  light beam perpendicular to incident direction gives rise to $100\%$ polarized radiation. This is why for $\mu\sim 0$ the small part of the  emerging radiation has large degree of polarization.
  
It should be mentioned that for sources of type $s(\tau)=Const$ and $s(\tau)\sim \exp(-h\tau)$ the peak like intensity does not hold
(see Silant'ev 1980). For these sources the increase of absorption $\varepsilon$ decreases the sharpness of angular dependence $J(\mu)$. As a result, the polarization of emerging radiation is small  compared with the generalized Milne problem.

In Table 1 we present the dependence of characteristic number $k$ on the degree of absorption $\varepsilon$, when the free
electrons are absent ($\eta=0$). It is seen that the value $k$ practically does not depend on the form of freely oriented dust particles.
The approximation value (31) also confirms this.
Nevertheless, the small inequalities $k_{sphere}<k_{plate}<k_{needle}$ exist.  More profound inequality occurs for needle like 
particles.  The relative difference ($\simeq 1\%)$ corresponds to  absorption factor $\varepsilon \simeq 
0.5$. For large absorption factor ($\varepsilon \to 1$) the $k$ - difference between all grain forms disappears. Our Table shows that the increase of parameter $\overline b_2 $ is accompanied by small increasing of parameter $k$. 

In Table 2 we demonstrate how the presence of the second component, i.e. the free non-absorbing electrons, affects the value of
characteristic number. Physically clear that for $\eta>1$  the mean effective absorption diminishes, i.e. the value $k$  also diminishes.
Tables 1,  2  and Fig.1 demonstrate this feature  for all range of parameter $\eta$ . 
 The large values $\eta$ corresponds to small $k$ (see Fig.1).  Recall, that Fig.1 corresponds to $\overline{b}_1=1$. Clearly, the small characteristic number does not give very sharp angular dependence of the radiation  intensity $J(\mu)$. The case $\eta\to \infty$ corresponds to the free electron atmosphere.

 The value $J(\mu=1)=3.06$ for $k=0$ and  spherical polarizability ($\overline{b}_1=1$) for non - absorbing particles is Chandrasekhar's (1960) case.  This case gives  the polarization $p(\mu=0)=11.71\%$.  In Table 3 we present also the Milne problem solutions for needle like particles ($\overline{b}_1=0.4$) and plate like ones ($\overline{b}_1=0.7$) . The needle like particles give $J(\mu=1)=2.96$ and corresponding maximum polarization $p(\mu=0)=3.80\%$ . For plate like grains we have $J(\mu=1)=3.00$ and $p(\mu=0)=7.26\%$. 

 For $\varepsilon=0.1$ and $\eta=0$ the corresponding results are given in Table 4. It is seen that the nonzero absorption increases
the polarization values for all forms of dust grains. The case $\eta\neq 0$ diminishes the polarization
degree, as it is seen from Table 5.

 We see that the presence of absorption ($\varepsilon=0.1$) considerably  increases the polarization values $p(\mu=0)=20.35\%, 5.48\%, 11.18\%$, for $\overline{b}_1=1,04,0.7 $, respectively. The angular dependence also increase, $J(\mu=1)=4.39, 4.02,$
$ 4.15$, respectively. For $\varepsilon=0.1$ and $\eta=1$  (see Fig. 5) the results  are: $p(\mu=0)=16.07\%$,

$ 8.45\%, 12.35\%$ and $J(\mu=1)=3.66, 3.46, 3.58$, respectively.

It is seen that anisotropy of particles considerably changes the polarization of emerging radiation. The
angular distribution $J(\mu)$ for all cases is near for $\overline{b}_1=1$ case.  Note that the polarization from
needle like particles is smaller than that from plate like ones.

 The limit case of $\varepsilon=1$  (pure absorbing dust grains) for $\eta=1,5$ and 10 is presented in Table 6. Note that for $\eta\to\infty$ the angular distribution and degree of polarization tend to values in Table 3 (two first columns).

 Figs. 2-5 present the  values of angular distribution of radiation $J(\mu)$ and polarization degrees $p(\mu)$ in \% for
 parameters $\eta=0,1,5, 10$, $\varepsilon=0.05,0.1,0.2,0.3$ and $\overline{b}_1=1, 0.4, 0.7$. It is seen that cases $\eta=5$ and $10$ are close to free electron scattering.
 
 Finally, we give short description of results in Figs. 2-5. First, the cases of spherical grains ($\overline{b}_1=1$ presented by the continuous curves)
have the greater $J(\mu)$ and $p(\mu)$ than those for  the anisotropic grains ($\overline{b}=0.7, 0.4$, the chain-dotted line and short-dotted  lines, respectively). The increase of the parameter $\eta$ diminishes the  polarization curves for $\overline{b}_1=1$ case.
In contrast, polarization curves for $\overline{b}_1=0.4$  increase with the increasing of $\eta$.  In the limit of great $\eta$
all curves tends to the electron scattering case with effective absorption $\varepsilon_{eff}=1/(1+\eta)$.

\begin{table}[h]
\caption { The characteristic number $ k$ for $\eta=0$ and $\overline{b}_1=1, 0.4, 0.7$ corresponding to spherical,
needle like and plate like particles, respectively}
\small
\begin{tabular}{| p{0.6cm} | p{1.3cm}p{1.3cm}p{1.3cm}p{1.3cm}}
\hline
$\varepsilon$ & $\overline{b}_1=1$ & $\overline{b}_1=0.4$ & $\overline{b}_1=0.7 $ & $k_{appr}$\\
\hline
0        & 0    &   0  &   0  & 0  \\
0.001 & 0.054743 & 0.054749  & 0.054747  &  0.05477\\
0.01   & 0.172285 & 0.172473  & 0.172417  &  0.17350\\
0.05   & 0.377166 & 0.379075  & 0.378489  &  0.38730\\
0.1     & 0.519583 & 0.524331 & 0.522813   &   0.54772\\
0.2     & 0.697604 & 0.707702 & 0.704204   &   0.77460  \\
0.3     & 0.811199 & 0.824504 & 0.819545  &   - \\
0.4     & 0.888707 & 0.902445 & 0.896992  &  -  \\
0.5     & 0.941298 & 0.952885 & 0.948054   &  -  \\
0.6     & 0.974750 & 0.982379 & 0.979104  &   -  \\
0.7     & 0.992822 & 0.996093 & 0.994698  &  -  \\
0.8     & 0.999292 & 0.999774 & 0.999588  &   -  \\
0.9     & 0.999999 & 0.999999 & 0.999999  &  -  \\
1        & 1               & 0.999999 & 0.999999  &  -   \\ 
\hline 
\end{tabular}
\end{table}
\normalsize

\begin{table}[h]
\caption {  The characteristic number $ k$ for $\eta=5$ and $\overline{b}_1=1, 0.4, 0.7$ corresponding to spherical,
needle like and plate like particles, respectively}
\small
\begin{tabular}{| p{0.6cm} | p{1.3cm}p{1.3cm}p{1.3cm}p{1.3cm}}
\hline
$\varepsilon$ & $\overline{b}_1=1.$ & $\overline{b}_1=0.4$ & $\overline{b}_1=0.7 $ &$k_{appr}$ \\
\hline
0        & 0    &   0  &   0  & 0 \\
0.001 & 0.022359 & 0.022359  & 0.022359 & 0.02236\\
0.01   & 0.070648 & 0.070652  & 0.070650 & 0.07071\\
0.05   & 0.157413 & 0.157458  & 0.157438 &  0.15811\\
0.1     & 0.221632 & 0.221750 & 0.221697  &  0.22361\\
0.2     & 0.310678 & 0.310969 & 0.310836 &  0.31623\\
0.3     & 0.377166 & 0.377625 & 0.377412 &  0.38730\\
0.4     & 0.431710 & 0.432302 & 0.432024 &   -  \\
0.5     & 0.478465 & 0.479138 & 0.478818  &  - \\
0.6     & 0.519582 & 0.520273 & 0.519941  &   - \\
0.7     & 0.549302 & 0.556990 & 0.556682  &  -  \\
0.8     & 0.589629 & 0.590134 & 0.589887 &   -  \\
0.9     & 0.620005 & 0.620297 & 0.620152 &   - \\
1        & 0.647915 & 0.647915 & 0.647915  &  -  \\
\hline 
\end{tabular}
\end{table}
\normalsize

\begin{table}[h!]
\caption {\small  The angular distribution $J(\mu)$ and degree of polarization $p(\mu)=-Q(\mu)/I(\mu)$ in \% for $\varepsilon=0$ and
$\eta=0$ for different forms of particles ($\overline{b}_1=1, 0.4, 0.7$). Here all the characteristic numbers $k$ are equal to zero. }
\small
\begin{tabular}{|p{0.5cm} | p{0.8cm} p{0.8cm}|p{0.8cm} p{0.6cm}| p{0.7cm} p{0.6cm}|}
\hline
$\mu$ & $J_{1}$ & $ p_{1}$ & $J_{0.4}$  &  $ p_{0.4}$ & $J_{0.7}$ & $p_{0.7}$     \\
\hline
0       & 1        & 11.713 &     1    & 3.80 &   1     &  7.26 \\
0.01 & 1.036 & 10.878 & 1.035 & 3.44 & 1.036 & 6.66    \\
0.02 & 1.066 & 10.295 & 1.064 & 3.20 & 1.065 &  6.25    \\
0.03 & 1.094 & 9.805  & 1.091 & 3.02 & 1.093 & 5.91     \\
0.04 & 1.120 & 9.374 & 1.116 &  2.85 & 1.118 & 5.61   \\
0.05 & 1.146 & 8.986 & 1.140 & 2.71 & 1.143 & 5.35    \\
0.06 & 1.170 & 8.631 & 1.164 & 2.57 & 1.167 & 5.11    \\
0.07 & 1.194 & 8.304 & 1.187 & 2.46 & 1.191 &  4.89    \\
0.08 & 1.218 & 8.000 & 1.210 & 2.35 & 1.214 & 4.69   \\
0.09 & 1.241 & 7.716 & 1.232 & 2.25 & 1.237 & 4.51  \\
0.10 & 1.264 & 7.449 & 1.254 & 2.16 & 1.259 & 4.34  \\
0.15 & 1.375 & 6.312 & 1.371 & 1.74 & 1.378 & 4.55    \\
0.20 & 1.483 & 5.410 & 1.472 & 1.46 & 1.482 & 3.01  \\
0.25 & 1.587 & 4.667 & 1.571 &1.24  & 1.584 & 2.57  \\
0.30 & 1.690 & 4.041 & 1.669 & 1.06 & 1.683 & 2.20  \\
0.35 & 1.791 & 3.502 & 1.765 & 0.90 & 1.782 & 1.89  \\
0.40 & 1.892 & 3.033 & 1.860 & 0.77 & 1.879 & 1.63  \\
0.45 & 1.991 & 2.619 & 1.954 & 0.66 & 1.975 & 1.40  \\
0.50 & 2.091 & 2.252 & 2.048 & 0.56 & 2.071 & 1.19  \\
0.55 & 2.189 & 1.923 & 2.141 & 0.48 & 2.167 & 1.01  \\
0.60 & 2.287 & 1.627 & 2.234 & 0.40 & 2.262 & 0.85  \\
0.65 & 2.385 & 1.358 & 2.326 & 0.33 & 2.356 & 0.71  \\
0.70 & 2.483 & 1.113 & 2.418 & 0.27 & 2.450 & 0.57  \\
0.75 & 2.580 & 0.888 & 2.510 & 0.21 & 2.544 & 0.46  \\
0.80 & 2.677 & 0.682 & 2.601 & 0.16 & 2.638 & 0.35  \\
0.85 & 2.774 & 0.492 & 2.692 & 0.11 & 2.731 & 0.25  \\
0.90 & 2.870 & 0.316 & 2.774 & 0.08 & 2.815 & 0.16  \\
0.91 & 2.890 & 0.282 & 2.793 & 0.07 & 2.834 & 0.15  \\
0.92 & 2.909 & 0.249 & 2.811 & 0.06 & 2.853 & 0.13  \\
0.93 & 2.928 & 0.216 & 2.829 & 0.05 & 2.871 & 0.11  \\
0.94 & 2.947 & 0.184 & 2.847 & 0.04 & 2.890 & 0.10  \\
0.95 & 2.967 & 0.152 & 2.865 & 0.04 & 2.909 & 0.08  \\
0.96 & 2.986 & 0.121 & 2.883 & 0.03 & 2.927 & 0.06  \\
0.97 & 3.005 & 0.090 & 2.902 & 0.02 & 2.946 & 0.05  \\
0.98 & 3.015 & 0.060 & 2.920 & 0.01 & 2.964 & 0.03  \\
0.99 & 3.044 & 0.030 & 2.938 & 0.01 & 2.983 & 0.02  \\
1    & 3.063 & 0     & 2.956 & 0    & 3.002 & 0     \\
\hline 
\end{tabular}
\end{table}
\normalsize

\begin{table}[h!]
\caption { \small The angular distribution $J(\mu)$ and degree of polarization $p(\mu)=-Q(\mu)/I(\mu)$ in \% for $\varepsilon=0.1$ and
$\eta=0$ for different forms of particles ($\overline{b}_1=1(k=0.519582), 0.4 (k=0.524331),0.7(k=0.522813)$)}.
\small
\begin{tabular}{|p{0.5cm} | p{0.8cm}p{0.8cm}| p{0.8cm}p{0.6cm}| p{0.7cm}p{0.6cm}|}
\hline
$\mu$ &  $J_{1}$  & $p_{1}$ & $J_{0.4}$  &  $p_{0.4}$  &  $J_{0.7}$  &  $p_{0.7}$  \\
\hline
0       & 1        & 20.348 &     1    & 5.48 &   1      & 11.19   \\
0.01 & 1.033 & 19.606 & 1.033 & 5.14& 1.033 & 10.62    \\
0.02 & 1.061 & 19.081 & 1.060 & 4.91 & 1.061 & 10.23    \\
0.03 & 1.087 & 18.636 & 1.085 & 4.73 & 1.086 & 9.91     \\
0.04 & 1.111 & 18.239 & 1.109 & 4.57 & 1.111 & 9.62   \\
0.05 & 1.135 & 17.879 & 1.133 & 4.42 & 1.135 & 9.37    \\
0.06 & 1.159 & 17.545 & 1.155 & 4.29 & 1.158 & 9.14    \\
0.07 & 1.182 & 17.233 & 1.178 & 4.18 & 1.181 &  8.92    \\
0.08 & 1.205 & 16.939 & 1.200 & 4.07& 1.203 & 8.73   \\
0.09 & 1.227 & 16.660 & 1.222 & 3.97& 1.225 & 8.54  \\
0.10 & 1.249 & 16.395 & 1.243 & 3.87 & 1.247 & 8.36  \\
0.15 & 1.371 & 15.099 & 1.360 & 3.44 & 1.366 & 7.54    \\
0.20 & 1.481 & 14.080 & 1.465 & 3.12& 1.474 & 6.93  \\
0.25 & 1.594 & 13.146 & 1.571 & 2.86 & 1.583 & 6.39 \\
0.30 & 1.710 & 12.264 & 1.680 & 2.62 & 1.694 & 5.91  \\
0.35 & 1.831 & 11.413 & 1.792 & 2.41 & 1.809 & 5.46  \\
0.40 & 1.957 & 10.576 & 1.908 & 2.21 & 1.929 & 5.02  \\
0.45 & 2.090 & 9.746 & 2.029 & 2.02 & 2.055 & 4.61  \\
0.50 & 2.231 & 8.914 & 2.156 & 1.83 & 2.187 & 4.20  \\
0.55 & 2.381 & 8.075 & 2.290 & 1.65 & 2.326 & 3.79  \\
0.60 & 2.541 & 7.226 & 2.431 & 1.47 & 2.475 & 3.39  \\
0.65 & 2.712 & 6.365 & 2.582 & 1.29 & 2.633 & 2.98  \\
0.70 & 2.897 & 5.489 & 2.744 & 1.11 & 2.802 & 2.57  \\
0.75 & 3.098 & 4.598 & 2.917 & 0.93 & 2.985 & 2.15  \\
0.80 & 3.316 & 3.691 & 3.105 & 0.74 & 3.183 & 1.73  \\
0.85 & 3.554 & 2.767 & 3.308 & 0.56 & 3.398 & 1.30  \\
0.90 & 3.789 & 1.923 & 3.507 & 0.39 & 3.609 & 0.90  \\
0.91 & 3.843 & 1.733 & 3.553 & 0.35 & 3.659 & 0.81  \\
0.92 & 3.899 & 1.543 & 3.601 & 0.31 & 3.709 & 0.72  \\
0.93 & 3.957 & 1.352 & 3.649 & 0.27 & 3.760 & 0.64  \\
0.94 & 4.015 & 1.161 & 3.698 & 0.24 & 3.812 & 0.55  \\
0.95 & 4.075 & 0.969 & 3.748 & 0.20 & 3.865 & 0.46  \\
0.96 & 4.135 & 0.776 & 3.799 & 0.16 & 3.920 & 0.36  \\
0.97 & 4.197 & 0.583 & 3.852 & 0.12 & 3.975 & 0.27  \\
0.98 & 4.261 & 0.389 & 3.905 & 0.08 & 4.032 & 0.18  \\
0.99 & 4.326 & 0.195 & 3.959 & 0.04 & 4.090 & 0.09  \\
1      & 4.392 & 0         & 4.015 & 0      & 4.149 & 0     \\
\hline 
\end{tabular}
\end{table}

\begin{table}[h!]
\caption {\small  The angular distribution $J(\mu)$  and degree of polarization $p(\mu)=-Q(\mu)/I(\mu)$ in \% for $\varepsilon=0.1$ and
$\eta=1$ for different forms of particles ($\overline{b}_1=1 (k=0.377166), 0.4 (k=0.378446), 0.7 (k=0.377953) $)}.
\small
\begin{tabular}{|p{0.5cm} | p{0.8cm}p{0.8cm}| p{0.8cm}p{0.6cm}|p{0.7cm}p{0.7cm}|}
\hline
$\mu$ & $J_{1}$ & $p_{1}$ &  $J_{0.4}$ & $p_{0.4}$ & $J_{0.7}$ & $p_{0.7}$ \\
\hline
0       & 1        & 16.073 &     1    & 8.45 &   1      & 12.35   \\
0.01 & 1.035 & 15.283& 1.035 & 7.86 & 1.035 & 11.65    \\
0.02 & 1.064 & 14.729 & 1.063 & 7.46 & 1.064 & 11.17    \\
0.03 & 1.904 & 14.261 & 1.089 & 7.14 & 1.090 & 10.76     \\
0.04 & 1.116 & 13.846 & 1.115 & 6.85 & 1.116 & 10.40  \\
0.05 & 1.141 & 13.472 & 1.139 & 6.60 & 1.140 & 10.08    \\
0.06 & 1.165 & 13.127 & 1.162 & 6.37 & 1.164 & 9.79    \\
0.07 & 1.188 & 12.807 & 1.185 & 6.16 & 1.187 & 9.52    \\
0.08 & 1.211 & 12.508 & 1.208 & 5.96 & 1.210 & 9.27   \\
0.09 & 1.234 & 12.227 & 1.230 & 5.78 & 1.233 & 9.03  \\
0.10 & 1.257 & 11.960 & 1.253 & 5.62 & 1.255 & 8.81  \\
0.15 & 1.379 & 10.694 & 1.371 & 4.84 & 1.376 & 7.78    \\
0.20 & 1.487 & 9.744 & 1.476 & 4.30 & 1.483 & 7.02  \\
0.25 & 1.596 & 8.914 & 1.580 & 3.84 & 1.590 & 6.37 \\
0.30 & 1.706 & 8.168 & 1.685 & 3.45 & 1.698 & 5.80  \\
0.35 & 1.817 & 7.479 & 1.791 & 3.11 & 1.807 & 5.28  \\
0.40 & 1.931 & 6.833 & 1.898 & 2.80 & 1.918 & 4.80  \\
0.45 & 2.048 & 6.216 & 2.007 & 2.52 & 2.032 & 4.35  \\
0.50 & 2.168 & 5.622 & 2.119 & 2.25 & 2.148 & 3.92  \\
0.55 & 2.291 & 5.042 & 2.234 & 2.00 & 2.268 & 3.50  \\
0.60 & 2.420 & 4.472 & 2.352 & 1.76 & 2.392 & 3.10  \\
0.65 & 2.553 & 3.908 & 2.473 & 1.53 & 2.521 & 2.70  \\
0.70 & 2.691 & 3.347 & 2.599 & 1.30 & 2.654 & 2.31  \\
0.75 & 2.836 & 2.784 & 2.730 & 1.08 & 2.793 & 1.92  \\
0.80 & 2.988 & 2.226 & 2.866 & 0.86 & 2.938 & 1.53  \\
0.85 & 3.146 & 1.662 & 3.008 & 0.64 & 3.090 & 1.14  \\
0.90 & 3.297 & 1.151 & 3.141 & 0.44 & 3.233 & 0.79  \\
0.91 & 3.331 & 1.037 & 3.171 & 0.40 & 3.265 & 0.71  \\
0.92 & 3.366 & 0.922 & 3.202 & 0.35 & 3.298 & 0.63  \\
0.93 & 3.401 & 0.808 & 3.233 & 0.31 & 3.331 & 0.55  \\
0.94 & 3.436 & 0.693 & 3.264 & 0.26 & 3.365 & 0.48  \\
0.95 & 3.472 & 0.578 & 3.295 & 0.22 & 3.399 & 0.40  \\
0.96 & 3.508 & 0.463 & 3.327 & 0.18 & 3.434 & 0.32  \\
0.97 & 3.545 & 0.348 & 3.359 & 0.13  & 3.468 & 0.24  \\
0.98 & 3.582 & 0.232 & 3.392 & 0.09 & 3.503 & 0.16  \\
0.99 & 3.619 & 0.116 & 3.425 & 0.04 & 3.539 & 0.08  \\
1      & 3.657 & 0         & 3.458 & 0      & 3.575 & 0     \\
\hline 
\end{tabular}
\end{table}

\begin{table}[h!]
\caption {\small  The angular distribution $J(\mu)$  and degree of polarization $p(\mu)=-Q(\mu)/I(\mu)$ in \% for limit case of pure absorbing   grains $\varepsilon=1$. The columns
correspond to $\eta=1 (k=0.94129807), 5 (k=0.64791542), 10 (k=0.49776253) $ }.
\small
\begin{tabular}{|p{0.5cm} | p{0.8cm}p{0.8cm}| p{0.8cm}p{0.6cm}|p{0.7cm}p{0.7cm}|}          
\hline
$\mu$ & $J_{1}$ & $p_{1}$ &  $J_{5}$ & $p_{5}$ & $J_{10}$ & $p_{10}$ \\
\hline
0       & 1        & 52.55 &     1    & 25.91 &   1      & 19.58   \\
0.01 & 1.022 & 52.19& 1.031 & 25.23 & 1.034 & 18.83    \\
0.02 & 1.042 & 51.91 & 1.058 & 24.75 & 1.062 & 18.30    \\
0.03 & 1.060 & 51.65 & 1.082 & 24.33 & 1.087 & 17.85     \\
0.04 & 1.079 & 51.40 & 1.106 & 23.96 & 1.112 & 17.45  \\
0.05 & 1.097 & 51.16 & 1.129 & 23.62 & 1.136 & 17.09    \\
0.06 & 1.116 & 50.91 & 1.151 & 23.30 & 1.160 & 16.75    \\
0.07 & 1.134 & 50.65 & 1.174 & 23.00 & 1.183 & 16.43    \\
0.08 & 1.153 & 50.40 & 1.196 & 22.72 & 1.206 & 16.14   \\
0.09 & 1.172 & 50.14 & 1.218 & 22.44 & 1.228 & 15.86  \\
0.10 & 1.191 & 49.87 & 1.239 & 22.18 & 1.251 & 15.59  \\
0.15 & 1.302 & 48.25 & 1.360 & 20.84 & 1.372 & 14.30    \\
0.20 & 1.414 & 46.54 & 1.472 & 19.72 & 1.482 & 13.30  \\
0.25 & 1.541 & 44.61 & 1.589 & 18.65 & 1.594 & 12.38 \\
0.30 & 1.687 & 42.45 & 1.713 & 17.58 & 1.710 & 11.53  \\
0.35 & 1.855 & 40.08 & 1.845 & 16.50 & 1.829 & 10.70  \\
0.40 & 2.051 & 37.51 & 1.987 & 15.40 & 1.953 & 9.90  \\
0.45 & 2.283 & 34.78 & 2.141 & 14.28 & 2.083 & 9.11  \\
0.50 & 2.561 & 31.89 & 2.309 & 13.12 & 2.220 & 8.32  \\
0.55 & 2.898 & 28.88 & 2.494 & 11.93 & 2.365 & 7.53  \\
0.60 & 3.315 & 25.77 & 2.698 & 10.71 & 2.319 & 6.73  \\
0.65 & 3.840 & 22.57 & 2.925 & 9.46 & 2.683 & 5.93  \\
0.70 & 4.520 & 19.32 & 3.180 & 8.17 & 2.859 & 5.11  \\
0.75 & 5.430 & 16.04 & 3.468 & 6.85 & 3.049 & 4.28  \\
0.80 & 6.706 & 12.74 & 3.797 & 5.50 & 3.254 & 3.43  \\
0.85 & 8.610 & 9.44  & 4.174 & 4.12  & 3.477 & 2.57  \\
0.90 & 11.342 & 6.48 & 4.565 & 2.86 & 3.694 & 1.79  \\
0.91 & 12.171 & 5.82 & 4.660 & 2.58 & 3.744 & 1.61  \\
0.92 & 13.116 & 5.17 & 4.757 & 2.30 & 3.796 & 1.43  \\
0.93 & 14.206 & 4.52 & 4.858 & 2.01 & 3.849 & 1.26  \\
0.94 & 15.475 & 3.87 & 4.963 & 1.73 & 3.902 & 1.08  \\
0.95 & 16.970 & 3.22 & 5.071 & 1.44 & 3.957 & 0.90  \\
0.96 & 18.759 & 2.57 & 5.183 & 1.15 & 4.012 & 0.72  \\
0.97 & 20.937 & 1.93 & 5.299 & 0.87  & 4.069 & 0.54  \\
0.98 & 23.645 & 1.28 & 5.419 & 0.58 & 4.127 & 0.36  \\
0.99 & 27.103 & 0.64 & 5.543 & 0.29 & 4.186 & 0.18  \\
1      & 3.657 & 0         & 3.458 & 0      & 3.575 & 0     \\
\hline 
\end{tabular}
\end{table}

\begin{figure*}[h!]
\fbox{\includegraphics[width=16cm, height=9cm]{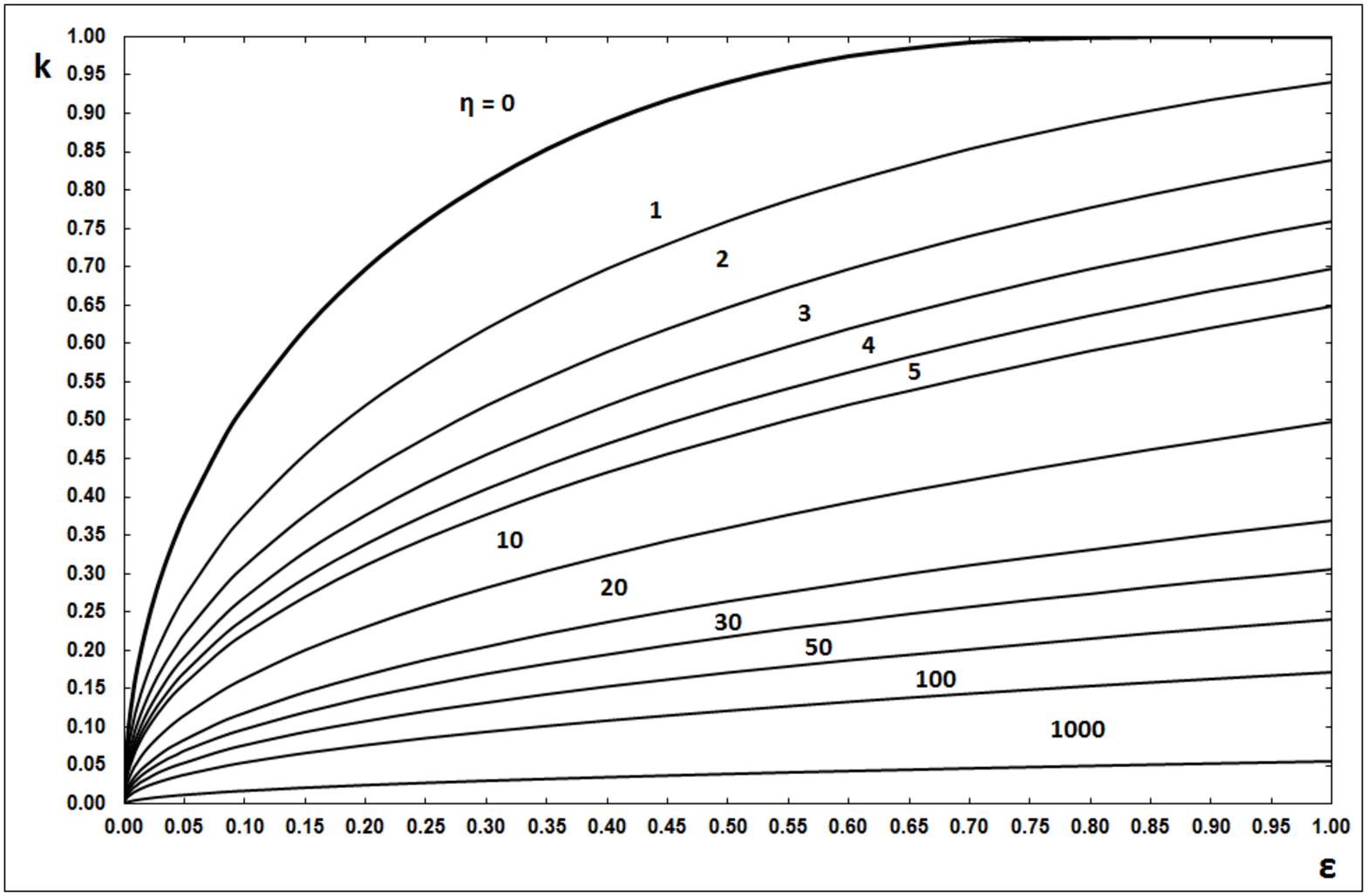}}
\caption{Dependence of characteristic number $k$ on degree of absorption $\varepsilon$ at $\overline{b}_1=1$.
The numbers denote the value of parameter $\eta$.}
\label{a}
\end{figure*}

\begin{figure*}[h!]
\fbox{\includegraphics[width=16cm, height=9cm]{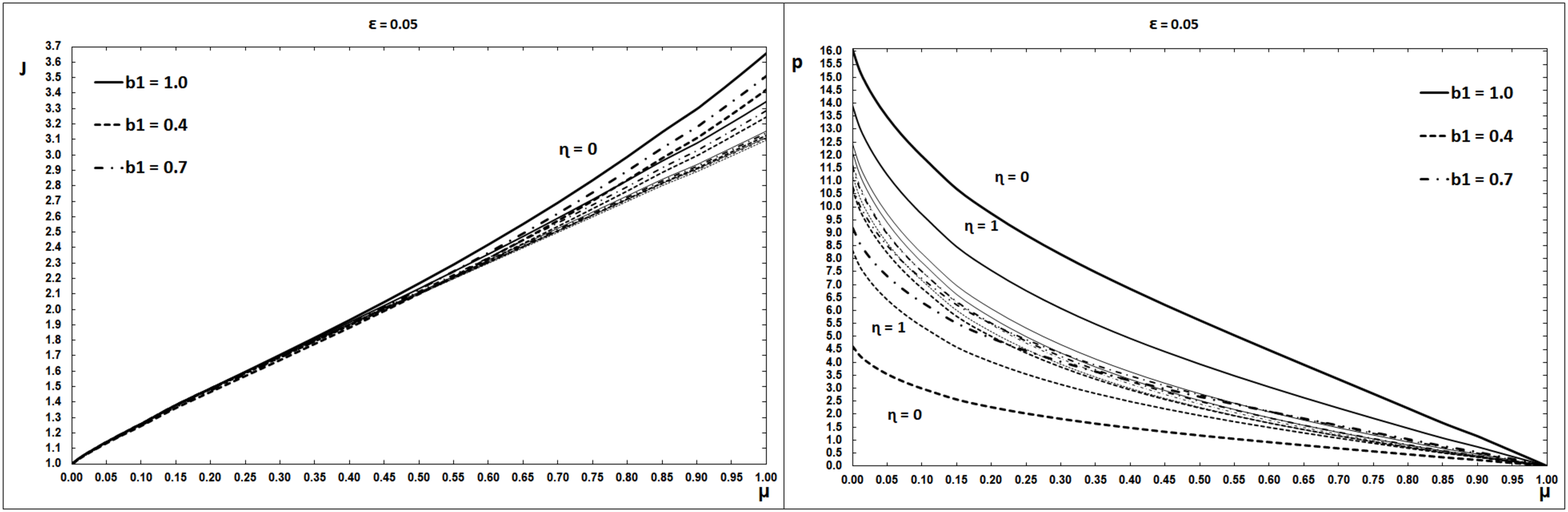}}
\caption{The angular distribution $J(\mu)$ and polarization degree $p(\mu)$ in \% for different forms of dust grains ($\overline{b}_1=1, 0.7, 0.4$ , describing by continuous, chain-dotted and short-dotted  lines, respectively ) and the values $\eta=0, 1,5, 10$.  The thicknesses of curves diminish corresponding  to values $\eta = 0,1,5,10$ .The degree of absorption $\varepsilon=0.05$.}
\label{a}
\end{figure*}

\begin{figure*}[h!]
\fbox{\includegraphics[width=16cm, height=9cm]{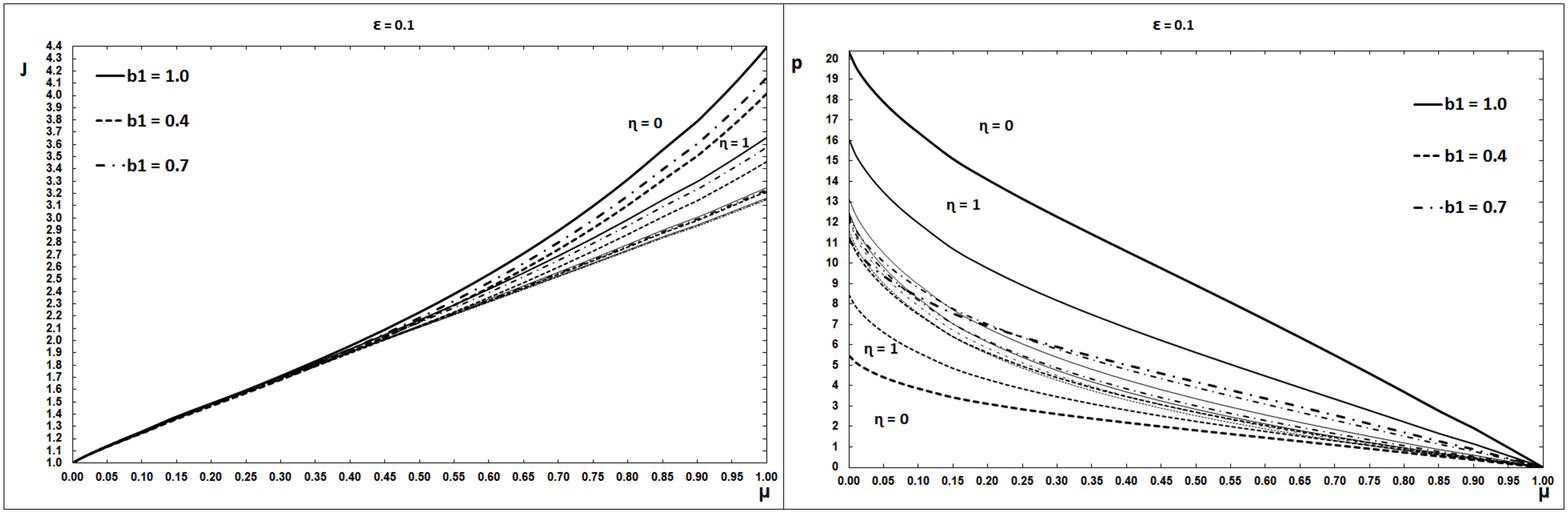}}
\caption{The angular distribution $J(\mu)$ and polarization degree $p(\mu)$ in \% for different forms of dust grains ($\overline{b}_1=1, 0.7, 0.4$,  describing by continuous, chain-dotted and short-dotted lines, respectively ) and the values $\eta=0, 1,5, 10$. The thicknesses  of curves diminish corresponding  to values $\eta=0,1,5,10$. The degree of absorption $\varepsilon=0.1$.} 
\label{a}
\end{figure*}

\begin{figure*}[h!]
\fbox{\includegraphics[width=16cm, height=9cm]{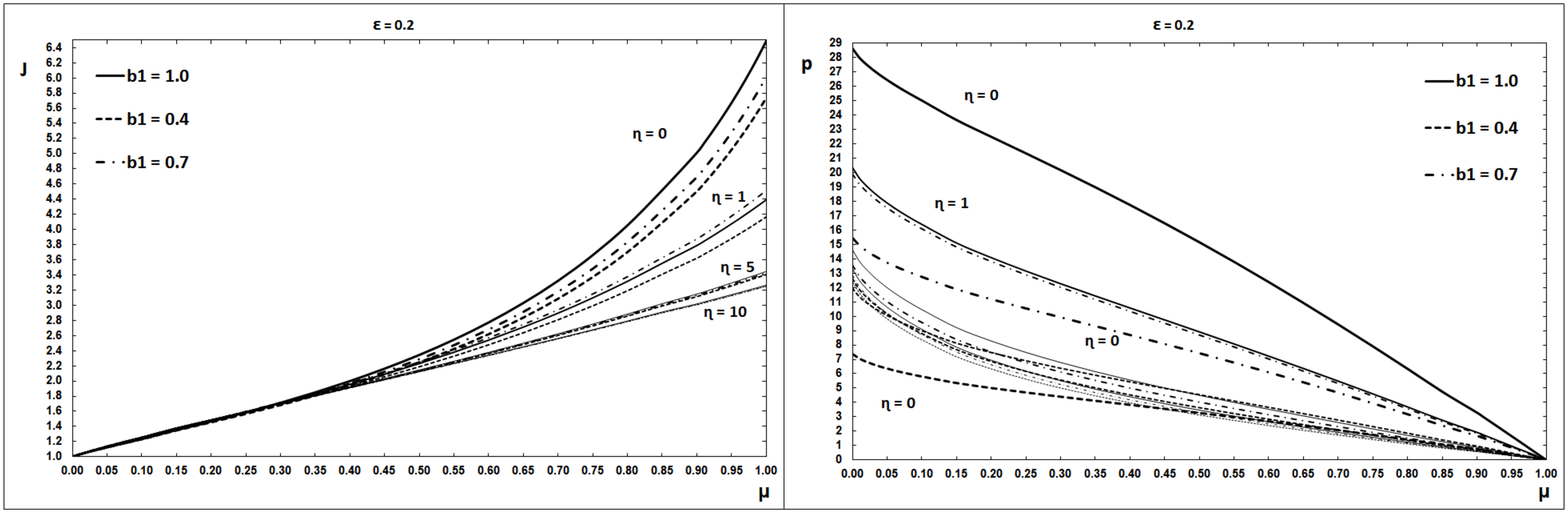}}
\caption{The angular distribution $J(\mu)$ and polarization degree $p(\mu)$ in \% for different forms of dust grains ($\overline{b}_1=1, 0.7, 0.4$,  describing by continuous, chain-dotted and short-dotted  lines, respectively ) and the values $\eta=0, 1,5, 10$. The thicknesses of curves diminish corresponding to values $\eta=0,1,5,10$.  The degree of absorption $\varepsilon=0.2$.}
\label{a}
\end{figure*}

\begin{figure*}[h!]
\fbox{\includegraphics[width=16cm, height=9cm]{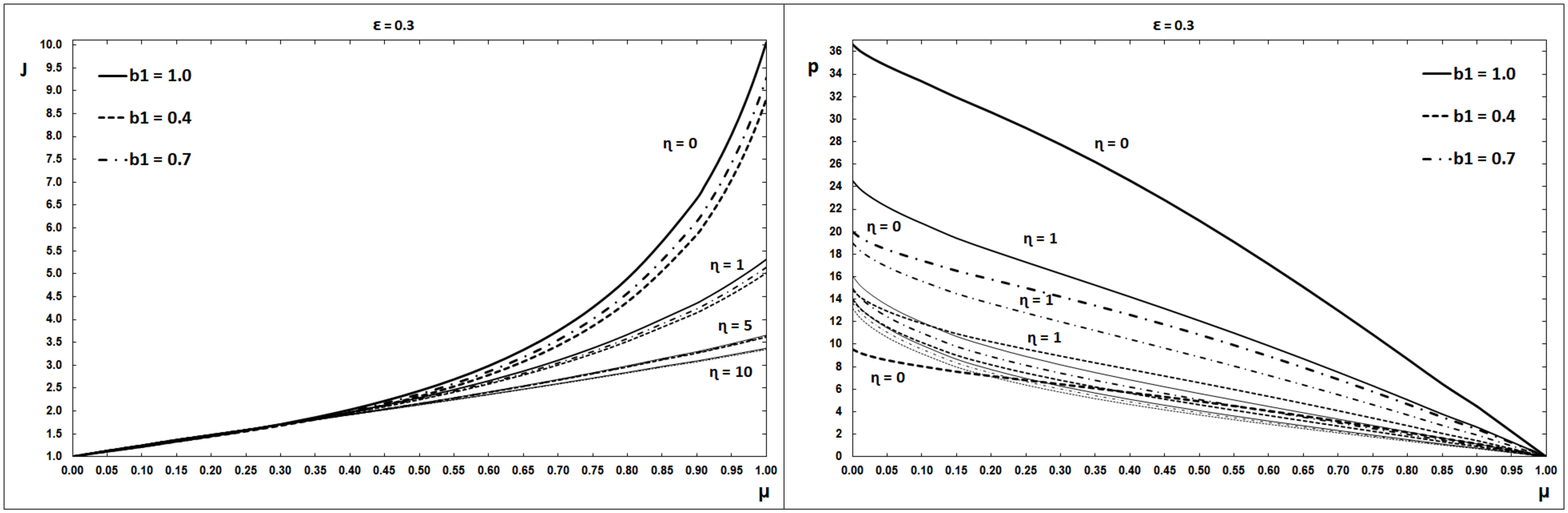}}
\caption{The angular distribution $J(\mu)$ and polarization degree $p(\mu)$ in \% for different forms of dust grains ($\overline{b}_1=1, 0.7 0.4$ , describing by continuous, chain-dotted and short-dotted  lines, respectively ) and the values $\eta=0, 1,5, 10$. The thicknesses of curves diminish corresponding to values  $\eta=0,1,5,10$. The degree of absorption $\varepsilon=0.3$.}
\label{a}
\end{figure*}

\section{Conclusion}

 We consider the generalized Milne problem in two components non-conservative atmosphere - free electrons and small absorbing anisotropic grains. The radiative transfer equation in such atmosphere depends on  two parameters $W$ and $(a+b)$, which depend on three physical
parameters. Parameter $\eta$ is the ratio of the Thomson optical length to
that due to small grains. The value $\overline{b}_1$  characterizes radiation scattered in accordance with Rayleigh phase matrix and
the parameter $\overline{b}_2$ (depolarization parameter) describes the isotropic scattering.  These values  obey the relation  $\overline{b}_1 +3 \overline{b}_2 =1$.  
The third parameter $\varepsilon=\sigma^{(a)}_{grain}/\sigma^{(t)}_{grain}$ is the degree of radiation absorption by dust grains.
 We  use the matrix resolvent function $\hat {R}(\tau,\tau')$  approach (see Silant'ev et al. 2015). This matrix is connected with two auxiliary matrices, depending on one variable - $R(\tau,0)$ and  $R(0,\tau')$. The Laplace transform of the matrix $R(\tau,0)$ with the parameter $1/\mu$  is the  matrix  $\hat H(\mu)$  function, which obeys  the nonlinear matrix equation  similar to known system of equations for Chandrasekhar's scalar $H$- functions. This matrix  obeys also the linear  integral equation with singular kernel.  This  linear equation was derived in algebraic manner from the  of nonlinear equation for matrix $\hat H(\mu)$ . Our calculations demonstrate that even small number of dust
grains in an atmosphere changes significantly the polarization of outgoing radiation as compared with  known Chandrasekhar's value
of polarization for free electron atmosphere.

{\bf Acknowledgements.} 
This research was supported by the Program of Prezidium of Russian Academy of Sciences N 17, by the Program of the Department of Physical Sciences of Russian
Academy of Sciences N 2 and the president program " the leading scientific schools" N 7241.

The authors are very grateful to Dr. H. Frisch for a number of useful remarks, especially for  new factorization (9).

\end{document}